# Boundary Guidance Hierarchical Network for Real-Time Tongue Segmentation

Xinyi Zeng, Qian Zhang, Jia Chen, Guixu Zhang, Aimin Zhou and Yiqin Wang

*Abstract*—Automated tongue image segmentation in tongue images is a challenging task for two reasons: 1) there are many pathological details on the tongue surface, which affect the extraction of the boundary; 2) the shapes of the tongues captured from various persons (with different diseases) are quite different. To deal with the challenge, a novel end-to-end Boundary Guidance Hierarchical Network (BGHNet) with a new hybrid loss is proposed in this paper. In the new approach, firstly Context Feature Encoder Module (CFEM) is built upon the bottom-up pathway to confront with the shrinkage of the receptive field. Secondly, a novel hierarchical recurrent feature fusion module (HRFFM) is adopt to progressively and hierarchically refine object maps to recover image details by integrating local context information. Finally, the proposed hybrid loss in a four hierarchy—pixel, patch, map and boundary guides the network to effectively segment the tongue regions and accurate tongue boundaries. BGHNet is applied to a set of tongue images. The experimental results suggest that the proposed approach can achieve the latest tongue segmentation performance. And in the meantime, the lightweight network contains only 15.45M parameters and performs only 11.22GFLOPS.

## I. INTRODUCTION

TONGUE is a unique vital organ. There are significant differences in tongue features even among twins. According to Traditional Chinese Medicine (TCM), the tongue has a certain distribution rule for internal organ lesions, such as spleen, stomach, hepatobiliary left, hepatobiliary right, kidney, and cardiopulmonary regions [1], as shown in Fig. 1. Therefore, the tongue is a potentially important body sign in clinical diagnosis and treatment, especially in TCM [2]. Doctors observe the shape and color of the tongue to judge human health or disease development [3], [4]. Tongue diagnosis is a noninvasive, effective method of auxiliary diagnosis anytime anywhere, which meets the needs of primary health care systems around the world. However, traditional tongue diagnosis has its limitations. First, the clinical competence of tongue diagnosis is determined by the experience, knowledge and thinking mode of the physicians. Second, environmental factors, such as differences in position and brightness of light, have an influence on the tongue diagnosis results. Finally, traditional tongue diagnosis is intimately related to the identification of diseases, and it is not well understood by western medicine and modern biomedicine. Moreover, making full use of medical resources to meet the needs of the patients always remains an urgent concern [5]. In order to avoid the malpractice of traditional tongue diagnosis, such as misdiagnosis, uncertainty and non-quantification, and to help people obtain cheap and high-quality medical services [6]-[10], it is necessary to build an objective, scientific and quantitative diagnostic standards for tongue diagnosis.

Therefore, in recent years, with the development of various technologies, image-based tongue diagnosis has attracted much attention. The tongue image shows the color, texture, and geometric features of the tongue, which can be used for automatic tongue diagnosis. The complete and accurate segmentation of tongue from tongue image plays a key role in tongue image automatic diagnosis, which directly affects the accuracy of tongue diagnosis [11], [12]. Nevertheless, the collected images not only show the tongue, but also teeth, lips, face and other materials, as shown in Fig. 2. Tongue appearance shows great differences among individuals, and the data is unbalanced, which make segmentation more difficult. For automatic disease diagnosis, it is very important to accurately segment the tongue area from the complex backgrounds.

A lot of efforts have been dedicated to tongue segmentation. Aplenty methods have been proposed in the past for the segmentation of tongue. Most of those are based on classical and commonly used image segmentation techniques. However, when the tongue area (the blue solid line shown in Fig. 3) is similar to the non-tongue area, some have poor segmentation [12]-[14], some are sensitive to noise or clustered backgrounds [15], [16], and some achieve relatively promising segmentation

Corresponding authors: Qian Zhang; Jia Chen.
    X. Zeng, Q. Zhang, G. Zhang, and A. Zhou are with the Shanghai Key Laboratory of Multidimensional Information Processing, School of Computer Science and Technology, East China Normal University, Shanghai 200062, China (e-mail: 51184506078@stu.ecnu.edu.cn; qzhang@cs.ecnu.edu.cn).
    J. Chen is with the Department of Integrative Medicine on Pediatrics, Shanghai Children's Medical Center, Shanghai Jiao Tong University School of Medicine, Shanghai 200127, China (e-mail: chenjia@scmc.com.cn).
    Y. Wang is with Shanghai Key Laboratory of Health Identification and Assessment and Laboratory of Traditional Chinese Medicine Four Diagnostic Information, Shanghai University of Traditional Chinese Medicine.

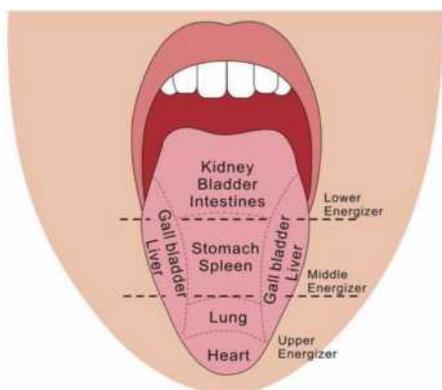

Fig. 1. Areas of the tongue corresponding to different internal organs.

results but their segmentation quality largely depends on some prior knowledge [17], [18]. Recently, convolutional neural networks (CNNs) have shown remarkable performance in tongue image segmentation [20], [21]. Although these CNNs-based methods achieve significant results compared with traditional methods, their tongue segmentations are still defective in fine structures and boundaries, as shown in Fig. 3. For the problem of object structures, most of semantic image methods use the U-shape structure to gradually increase the spatial resolution and fill some missing details [21]-[23]. Although these methods have achieved good performance, there are still many areas to be improved. For example, [22] has expensive computation with 30× more parameters than ENet [21], which drops the last stage of the model in pursuit of an extremely tight framework. PSPNet [23] and Deeplab v2 [25] use the dilated convolution to enlarge the receptive field, which preserves the spatial size of the feature map. Besides, most methods focus on region accuracy and ignore the boundary accuracy of the object. In order to get precise boundaries, some methods employ time-consuming CRF (Conditional Random Field) to refine the final predicted maps [24], [25]. Some methods specifically formulate the problem statement as binary or category-aware semantic edge detection [26]-[29], [48]. However, these methods did not monitor the boundaries of the final segmentation result of the network. They train the object boundary detection and object segmentation separately. For example, Discriminative Feature Network (DFN) [29] contains Smooth Network and Border Network. Smooth Network handles the object segmentation issue and Border Network handles the boundary detection issue. DFN's boundary detection can only be used as an auxiliary training for object segmentation, but cannot make the segmented object boundary be significantly improved.

Based on the above analysis, a novel Boundary Guidance Hierarchical Network (BGHNet) is proposed for real-time tongue segmentation in this paper. BGHNet contains two parts: a Context Feature Encoder Module (CFEM) and a novel Hierarchical Feature Refinement Module (HFRM). The CFEM is devised to confront with the shrinkage of the receptive field, and the HFRM hierarchically and progressively object maps to recover image details respectively. For CFEM, a lightweight CNN is adopted for feature extraction, and then several average pooling layers are appended on top of the backbone to capture global guidance information (where the tongue is). In the end, different levels of global average-pooled features are extracted to compute channel-wise attention at different level feature maps. For HFRM, the fusion of global (coarse) and local (fine) contexts as well as the refinement of object maps are investigated to achieve better accuracy without loss of speed. The Hierarchical recurrent connections are incorporated into each layer to generate high-resolution object maps. Inspired by [32], [42], this paper also proposes a hybrid loss, which combines four parts, namely, Binary Cross Entropy (BCE) [36], Structural SIMilarity (SSIM) [45], F1-score(F1) [31] and Boundary F1-score (BF1) [32] loss. The network is guided by the proposed hybrid loss to learn from ground truth in four hierarchies—pixel, patch, map and boundary levels. A series of ablation experiments are designed to allow readers to better understand the impact of each component in the proposed architecture, and to show how joint training with boundary detection can help to enhance boundary details of predicted results. To the best of our knowledge, such lightweight and boundary-quality network have not been used in medical image segmentation field yet.

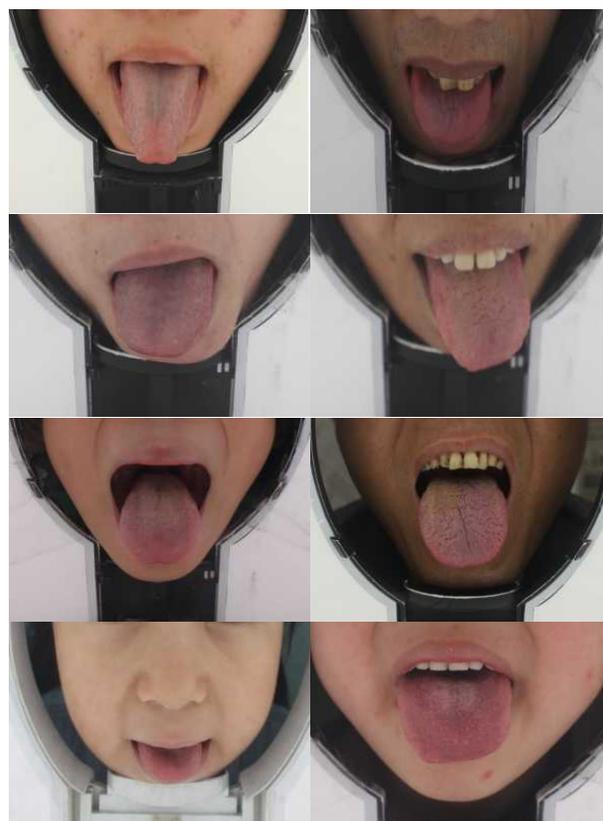

Fig. 2. Examples of tongue images with large variations of tongue appearances from different patients.

The remainder of this paper is organized as follows. The details of the proposed method are introduced in Section II. Experiments and results are reported in Section III. Conclusions are drawn in Section IV.

## II. METHODS

The proposed BGHNet consists of two modules: the Context Feature Encoder Module (CFEM) and the Hierarchical Feature

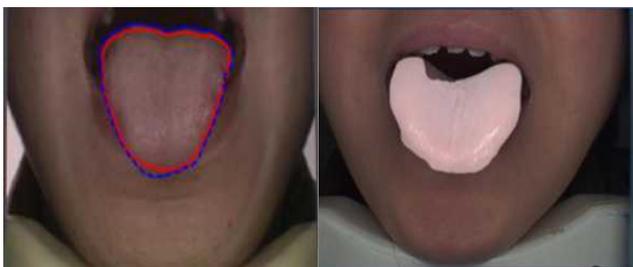

Fig. 3. Sample segmentation results of CNNs. (a) Red solid line indicates the prediction of [19] and Blue dashed line means the ground truth. (b) White transparent area indicates the prediction of [20].

Refinement Module (HFRM), as shown in Fig. 4. The CFEM first coarsely detects the target in a global perspective, then the HFRM hierarchically and progressively refines the details of the object map step by step. BGHNet is trained end-to-end. In the test, the predicted segmentation results can be concluded through the proposed network, without using any post-processing and other methods. Therefore, BGHNet is not only lightweight but also efficient.

### A. Context Feature Encoder Module

The CFEM is designed to provide sufficient receptive field. In the semantic segmentation task, the receptive field is of great significance for the performance [33]. As the receptive field expands, the position of the target becomes more and more precise. However, at the same time, spatial details are also ignored. Therefore, different levels of feature maps extracted in the CFEM are utilized in the HFRM to refine the high-level features with the local information. Since the contributions of different level local spatial information to high level feature information are different, a specific Global refinement block (GRB) [34], [35], which is an improved version of SE block [34], is proposed to refine features at various stages. The specific structure is shown in Fig. 5(a). Its role at different layers is different throughout the network. GRB can learn to use global information to selectively filter informative features. In shallower layers, it stimulates informative features in a class-agnostic way, reinforcing the shared low-level representations. In deeper layers, it becomes more specialized and responds to different inputs in a highly class-specific way. This is a top-down structure. The top-down pathway of BGHNet is built upon the bottom-up light backbone. However, there is a problem with this structure that the high-level features will be gradually diluted when they are transmitted to lower layers. [23] shows that the theoretical receptive fields of CNNs are far bigger than theoretical receiving fields, particularly for deeper layers. Therefore, the receptive fields of the whole networks are smaller and the global information of the input image cannot be captured. Regarding the lack of high-level semantic information for fine-level feature maps in the top-down pathway, a context Guidance Module (CGM) containing a modified version of the Pyramid Pool Model (PPM) [23] and the Global Feature Pyramid Model (GFPM) was proposed to accurately capture the precise position of the highlighted objects.

**Light Backbone:** the proposed architecture consists of downsampled-split-shuffle-non-bottleneck unit (DSS-nbt) and split-shuffle-non-bottleneck (SS-nbt). The structure of DSS-nbt is shown in Fig. 5(d). Inspired by [43], downsampled unit and split-shuffle-non-bottleneck (SS-nbt) are employed in the design of DSS-nbt, which is approaching to the representational power of large and dense layers, but the computational complexity is greatly reduced. Each DSS-nbt with stride 2 enables more deeper network connection context and reduce computation simultaneously. There are four stages in total. Except for the second stage that is composed of DSS-nbt and two SS-nbt, each stage is composed of one DSS-nbt and three SS-nbt. Moreover, the usage of dilated convolutions [44] in the last two stages of SS-nbt can effectively reduce the parameters and computational cost compared to the use of larger kernel sizes [43].

**Global Refinement Block:** Global Refinement Block (GRB) is designed to change the weights of feature maps of each stage in the feature network, as illustrated in Fig. 5(a). The first component of GRB is a global average pooling layer to capture global context, then the following is a 1 × 1 convolution layer. The number of channels is set according to the number of channels in the feature of each layer. GRB can combine the information across all global context. Finally, there is an activation function. GRB can learn to use global information to selectively stress informative features and suppress relatively useless ones.

**Context Guidance Module:** Context Guidance Module (CGM) is composed of PPM, GFPM and Sliced Concatenation (SC). SC is a modified version of Sliced Concatenation [27], reducing parameters and increasing information exchange. The CGM structure is shown in Fig. 5(c). The PPM consists of four sub-branches to capture the context information of the input images. The first and second sub-branches are the identity maps selected by the GRB layer and a global average pooling layer, respectively. For the last two middle sub-branches, adaptive average pooling layer are used to ensure the spatial sizes of output feature maps are 3×3 and 5×5, respectively. In a way to generate a feature pyramid containing global information, shuffle concatenation is used for information fusion, and group convolution is used to reduce the computation of concatenation. The flow of information between channel shuffle is caused by group convolution, so channel shuffle is used to enhance the ability of information representation. The modified SC structure is shown in Fig. 5(b). The Context Feature Pyramid Module (CFPM) consists of four sub-branches that are delivered to feature maps at various levels. Every branch adopts convolution layer and efficient sub-pixel convolution layer [38] to obtain feature maps of global information at different scales, such as 32, 64, 128 and 256.

### B. Hierarchical Feature Refinement Module

To further combine these features effectively and improve predicted object maps in details, a novel Hierarchical Feature Refinement Module is proposed to progressively and hierarchically render image details by integrating local context information. The core of the HFRM is the Feature Fusion Amplifier Block (FFAB). FFAB incorporates hierarchical connections into each layer and generates high-resolution object maps. As shown in Fig. 4, The first FFAB combines feature maps of 32x32 in CFPM with the fourth layer of the same scale feature maps in the backbone to generate object maps of 64x64. The second FFAB combines object maps of 64x64 in the first FFAB with feature map of 64x64 in CFPM and the third layer of the same scale feature maps in the backbone generates object maps of 128x128 (the subsequent further refined object maps are denoted in the same way). In Fig. 4, the detailed framework of a refinement step is shown. FFAB is adopted to combine the coarse object maps, feature maps of global information and the local features in the backbone, and enlarge the object maps one by one. HFRM refines the object maps in a coarse to fine and global to local way.

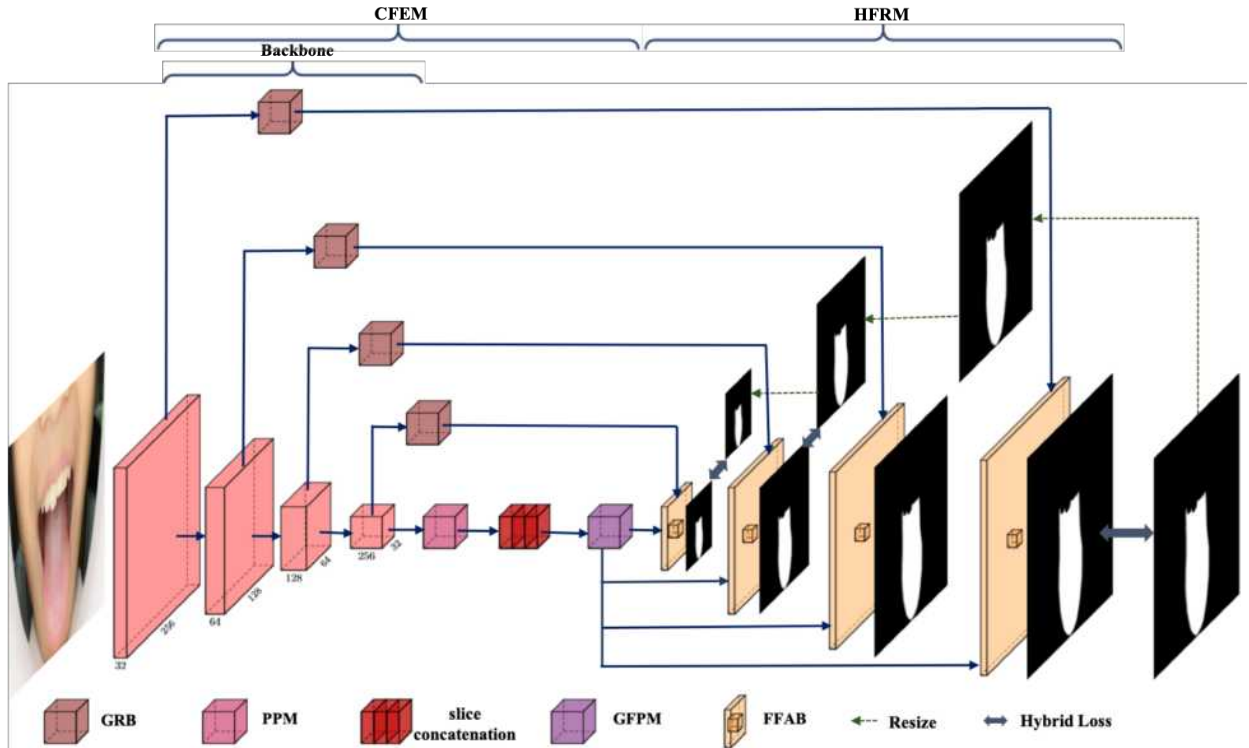

Fig. 4. The architecture of the proposed BGHNet method. The length of cuboid indicates the spatial size, while the thickness represents the number of channels. The different level feature maps in the CFEM utilized in the HFRM are shown. Firstly, the images are fed into Context Feature Encoder Module (CFEM). The CFEM is proposed to generate more high-level semantic feature maps. It contains a light backbone, four Global Refinement Block (GRB), a Pyramid Pool Model (PPM), a Sliced Concatenation (SC) and a Global Feature Pyramid Model (GFPM). Finally, the extracted features are fed into the Hierarchical Feature Refinement Module (HFRM) to obtain the mask as the segmentation prediction map. The HFRM contains four Feature Fusion Amplifier Block (FFAB) to enlarge the feature size, which replaces the original up-sampling operation.

**Feature Fusion Amplifier Block:** features at different levels convey different information, so these features are not simply summed up. The spatial information captured by the backbone encodes most rich detail information. Moreover, the output feature of the Context Feature Pyramid Module mainly encodes context information. Therefore, given the different level of the features, Feature Fusion Amplifier Block (FFAB) is designed to combine the feature information effectively. After effective fusion of features, the fused feature map needs to be decoded into object maps. Decoder can recover the pixelwise prediction from the lower-resolution feature maps. In previous works [39]-[41], decoder generally consists of a few deconvolutional layers or bilinear upsampling. However, deconvolutional has too many parameters to train. And, Bilinear upsampling has limited capability in recovering the pixel-wise prediction accurately, due to it does not consider the correlation among the prediction of each pixel.

In a way to solve the above upsampling issue, the proposed FFAB also contains a simple and effective data-dependent upsampling method, namely PixelShuffle [38], which can recover the pixelwise segmentation prediction from lower-resolution feature maps. PixelShuffle is originally proposed to solve the problem of image super resolution [38]. The main function of PixelShuffle is to obtain high resolution feature maps through convolution and multi-channel recombination of low-resolution object maps. Instead of directly generating the high-resolution object maps through interpolation, the process first obtains the feature maps of $r^2$ channels through convolution, and then obtains the object maps through periodic shuffling, where r is the upscaling factor, that is the enlargement ratio of the object maps, as is shown in Fig. 6(b). No special coding is required, and the resolution of the fused low-level features can be decoupled from the final predicted resolution. This decoupling extends the design space of the decoder's feature aggregation so that arbitrary feature aggregation can be used to improve the segmentation performance as much as possible.

### C. Hybrid Loss

The training loss is defined as follows:

$$\text{Loss} = \sum_{k=1}^{N} loss_T^{(k)} \qquad (1)$$

where, $loss^{(k)}$ is the loss of the $k$-th output, N the outputs' total number. The different losses are used to supervise the different output. As described above, our model has four outputs, three for sides and one for final. T stands for side or final output. T stands for side or final output.

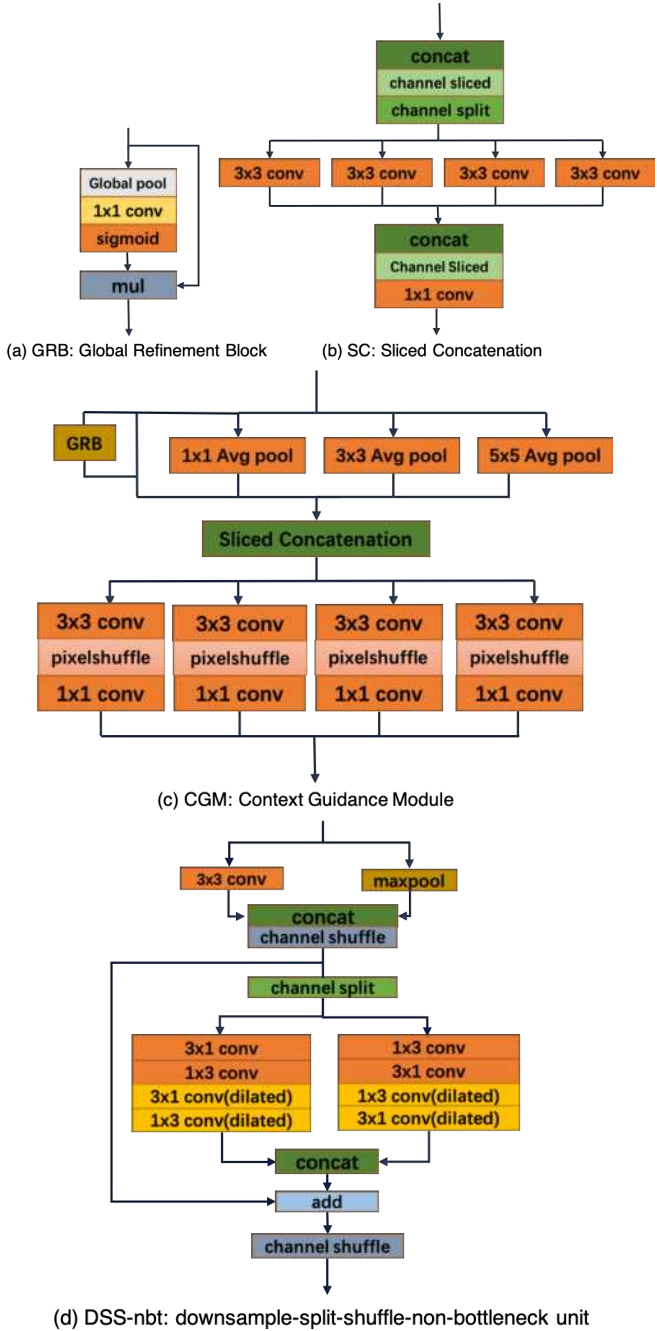

Fig. 5. The component of the Context Feature Encoder Module

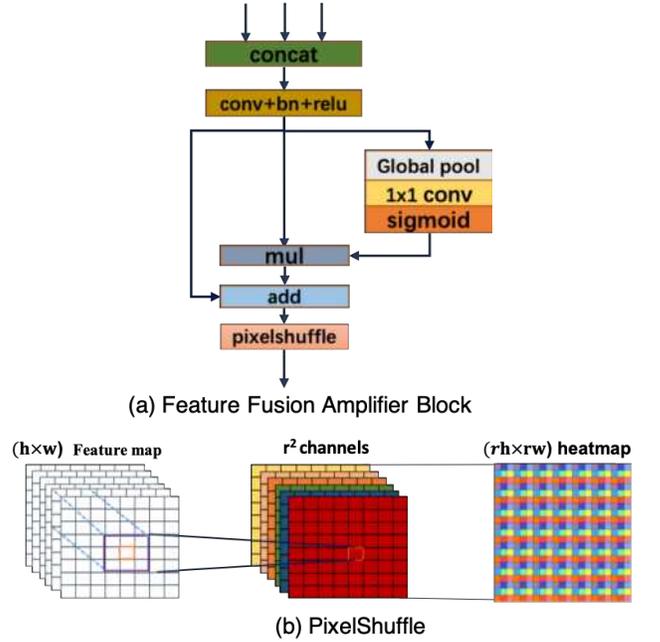

Fig. 6. The component of the Hierarchical Feature Refinement Module

The side loss of hybrid loss is designed as equation (2) to obtain high quality of region segmentation and accurate boundaries:

$$loss_{side}^{(k)} = loss_{bce}^{(k)} + loss_{F1}^{(k)} + loss_{BF1}^{(k)} \quad (2)$$

where $loss_{bce}^{(k)}, loss_{F1}^{(k)}, loss_{BF1}^{(k)}$ denote BCE loss [36], F1 loss [31], BF1 loss [32], respectively.

Although the loss function of side output can better explicitly supervise the training of the area and boundary, the boundary obtained by directly using the loss function of side output is more concave and convex. Therefore, the hybrid loss for the final outputs use SSIM which can focus on both boundary accuracy and regional accuracy to obtain a smoother and more detailed boundary, as defined below:

$$loss_{final}^{(k)} = loss_{bce}^{(k)} + loss_{F1}^{(k)} + loss_{ssim}^{(k)} \quad (3)$$

where $loss_{ssim}^{(k)}$ denote SSIM loss [45].

BCE loss [36] is defined as:

$$loss_{BCE} = \sum_{(i,j)} |G(i,j)\log(P(i,j)) + (1 - G(i,j))\log(1 - P(i,j))| \quad (4)$$

where for pixel(i, j) the ground truth label G(i, j) is in the {0, 1} and P(i, j) is the predicted probability of target object.

F1-score is a measure usually used to classify problems. It is the average of the accuracy rate and the recall rate. F1 loss is defined as:

$$loss_{F1} = 1 - \frac{2 \times P \times R}{P + R} \quad (5)$$

Here *P* refers to the ratio of positive examples to the total number of positive examples that the pixel is predicted. R is the ratio of positive examples to the total number of positive examples that the pixel is ground truth label. *P* and *R* are defined as:

$$P = \frac{sum(Pt \circ Gt)}{sum(Pt)}$$
$$R = \frac{sum(Pt \circ Gt)}{sum(Gt)} \quad (6)$$

where, ∘ denotes pixel-wise multiplication of two binary maps, and sum (·) summation of a binary map, $Pt \in \{0, 1\}$ and $Gt \in \{0, 1\}$ denote the binary map of predicted and ground truth for an arbitrary class c of one image, respectively.

The binary boundary mapping is needed to be extracted in order to construct various of BF1. The definition of boundary is as follows:

$$\begin{aligned} Pt^b &= pool(1 - Pt^b, \theta) - (1 - Pt^b) \\ Gt^b &= pool(1 - Gt^b, \theta) - (1 - Gt^b) \end{aligned} \quad (7)$$

where pool (·,·) denotes pixel-wise max-pooling operation with a sliding window of size θ. The boundary width is $\lfloor \frac{\theta}{2} \rfloor$, and θ is usually set to 3. The extended boundary map should be obtained in order to compute Euclidean Distances from pixels to boundary, and it is defined as:

$$\begin{aligned} Pt^{b,ext} &= pool(Pt^{b,ext}, \theta') \\ Gt^{b,ext} &= pool(Gt^{b,ext}, \theta') \end{aligned} \quad (8)$$

The value of hyperparameter $\theta'$ can be set as not greater than the minimum distance between neighboring segments of the binary ground truth map. After that $P_c^b$ and $R_c^b$ can be calculated as follows:

$$\begin{aligned} P^b &= \frac{sum(Pt^b \circ Gt^{b,ext})}{sum(Pt^b)} \\ R^b &= \frac{sum(Pt^{b,ext} \circ Gt^b)}{sum(Gt^b)} \end{aligned} \quad (9)$$

Finally, the boundary loss function is defined as:

$$loss_{BF1} = 1 - \frac{1}{C+1} \sum_{c=0}^{C} \frac{2 \times P^b \times R^b}{P^b + R^b} \quad (10)$$

SSIM [45] can capture the structural information in images. Given two images Pt and Gt, the structural similarity loss of predicted binary map and ground truth label can be calculated as follows:

$$loss_{SSIM} = 1 - \frac{(2\mu_{Pt}\mu_{Gt} + c_1)(2\sigma_{PtGt} + c_2)}{(\mu_{Pt}^2 + \mu_{Gt}^2 + c_1)(\sigma_{Pt}^2 + \sigma_{Gt}^2 + c_2)} \quad (11)$$

where $\mu_{Pt}$ and $\mu_{Gt}$ are the mean of Pt and Gt, respectively, $\sigma_{Pt}^2$ and $\sigma_{Gt}^2$ are the variance of Pt and Gt, respectively, $\sigma_{PtGt}$ is the covariance of Pt and Gt, $c_1$ and $c_2$ are generally equal to $0.01^2$ and $0.03^2$, in order to avoid dividing by zero.

The visual impact of SSIM losses in image segmentation is illustrated. $loss_{side}^{(k)}$ is used for final outputs, the boundary of the target is more concave and convex, as shown in Fig. 7(b). $loss_{final}^{(k)}$ is used for all outputs, the boundary of the target is smooth and fine, as shown in Fig. 7(c).

BCE loss is pixel-level measure, which considers each pixel and maintains smooth gradient for all pixels. SSIM loss can measure patch-level by consider each pixel's local neighborhood, and assigns higher weights to the boundary, even when the boundary predicted probabilities are the same with the rest of foreground. BF1 loss is a boundary-level measure, which keeps attention mainly on a boundary and can better handles edge effects. F1 loss is a map-level, which gives more focus on the foreground. As the confidence of the network prediction of the foreground grows, the loss of the foreground reduces eventually to zero.

## III. EXPERIMENT

### A. Datasets and Implementation details

To verify the proposed method, two datasets are used for experiments. Segmentation experiments are conducted on two tongue datasets to evaluate the performance of our proposed BGHNet.

**1) Tongue Datasets:** We use two datasets in the experiments. The first one, called Dataset1, is from an open tongue dataset, which contains 300 images with the size published by BioHit [49]. The second one, called Dataset2, is a self-built tongue, which is collected by us from the hospital and contains 1538 images with different sizes of 2304×3456, 1419×777 and so on. Tongue images in this Dataset2 were captured from thousands of patients, most of whom had large differences in color, texture and geometric features. Due to the different postures of patients, the shooting angles are different. Some tongue photos are positive, and some are on the side. Therefore, rotation data enhancement is performed for Dataset2 to train a more robust model. The number of datasets has been expanded to 4,614 images. We use 80% of the patients in Dataset1 and Dataset2 for training and the rest for testing. During training, after the shortest side of each image is scaled up to 512, the longest side is scaled by a multiple of the shortest side and randomly cropped to 512×512.

**2) Implementation Details:** the network is trained by using the mini-batch stochastic gradient descent (SGD) [46] with batch size 12, momentum 0.9 and weight decay 0.0001. Inspired by [25], [46], we use the "poly" learning rate policy is used, where the learning rate is multiplied by $(1 - \frac{iter}{max\_iter})^{power}$ with power = 0.9 and initial learning rate 0.01. The proposed network is implemented in Pytorch. All the weights of the convolution layers are initialized randomly. The validation dataset does not use during training. A GTX 2080 GPU is used for both training and testing.

### B. Evaluation Metrics

To evaluate algorithms' performance, three widely used and standard metrics are adopted, namely, mean intersection-over-union (MIoU), pixel-accuracy (PA) and F1-socre(F1). In addition, BF1 is used to evaluate target boundary.

**1) Mean Intersection over Union (MIoU):** it calculates a ratio between the intersection and the union of two sets, which in our case is the ground truth and predicted segmentation.

$$\text{MIoU} = \frac{TP}{(FP + FN + TP)} \quad (12)$$

**2) Pixel Accuracy (PA):** it simply calculates a ratio between the number of correctly classified pixels and the number of pixels in the entire image.

TABLE I
ABLATION ANALYSIS FOR THE PROPOSED ARCHITECTURE ON DATASET2

| No | Architecture | | | Dataset2 | | | | | |
|---|---|---|---|---|---|---|---|---|---|
| | GRM | CGM | FFAB | MIoU | PA | F1 | BF1 | | |
| | | | | | | | $\theta=3, \theta'=3$ | $\theta=3, \theta'=5$ | $\theta=3, \theta'=7$ |
| 1 | ✓ | ✓ | ✓ | 0.9911 | 0.9985 | 0.9955 | 0.9391 | 0.9758 | 0.9860 |
| 2 | | ✓ | ✓ | 0.9899 | 0.9984 | 0.9952 | 0.9402 | 0.9768 | 0.9866 |
| 3 | ✓ | | ✓ | 0.9906 | 0.9984 | 0.9952 | 0.9415 | 0.9774 | 0.9870 |
| 4 | ✓ | ✓ | | 0.9901 | 0.9983 | 0.9950 | 0.9233 | 0.9741 | 0.9856 |

TABLE II
ABLATION ANALYSIS FOR THE PROPOSED LOSS ON DATASET2

| No | Loss | Dataset2 | | | | | |
|---|---|---|---|---|---|---|---|
| | | MIoU | PA | F1 | BF1 | | |
| | | | | | $\theta=3, \theta'=3$ | $\theta=3, \theta'=5$ | $\theta=3, \theta'=7$ |
| 1 | $loss_{all} = \text{BCE}$ | 0.9863 | 0.9977 | 0.9930 | 0.5819 | 0.6796 | 0.7149 |
| 2 | $loss_{all} = BCE + F1 + BF1$ | 0.9908 | 0.9985 | 0.9954 | 0.9412 | 0.9772 | 0.9870 |
| 3 | $loss_{side} = BCE + F1 + BF1$, $loss_{final} = BCE + F1 + BF1 + SSIM$ | 0.9911 | 0.9985 | 0.9955 | 0.9391 | 0.9758 | 0.9860 |

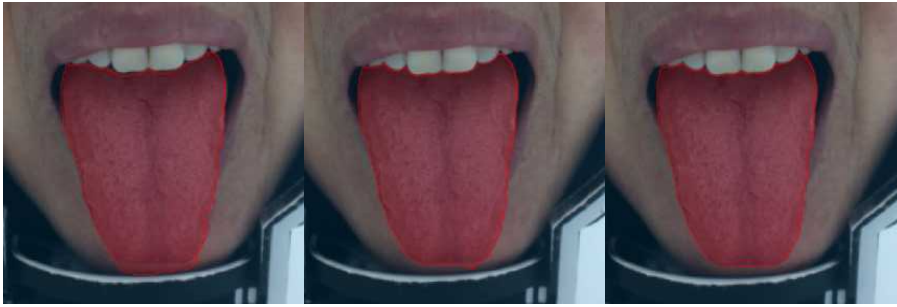

(a) The use of BCE loss for all output  (b) The use of $loss_{side}$ for final outputs.  (c) The use of $loss_{final}$ for final outputs

Fig. 7. The visual examples of tongue segmentation (red transparent area indicates the mask as the segmentation prediction map).

$$PA = \frac{TP + TN}{TP + TN + FP + FN} \quad (13)$$

where TP, FP, TN and FN represent the number of true positives, false positives, true negatives and false negatives.

3) **F1-score:** it comprehensively considers region precision and region recall.

$$F1 = \frac{2 \times P \times R}{P + R} \quad (14)$$

4) **BF1-score:** it comprehensively considers boundary precision and boundary recall.

$$BF1 = \frac{2 \times P^b \times R^b}{P^b + R^b} \quad (15)$$

*C. Ablation study*

In this section, the effectiveness of each key components used in our model is validated. The ablation study contains two parts: architecture ablation and loss ablation.

1) **Architecture ablation:** to verify the effectiveness of BGHNet, the quantitative comparison results of our model against other related architectures are reported. Except for different combinations of GRM, CGM and FFAB, all the other configurations are the same. TABLE I shows the performance on Dataset2.

● **Effectiveness of GRM:** in an effort to demonstrate the effectiveness of GRM, the GRM at different levels is removed. The results are listed in the 2nd row of TABLE I. There is a significant decrease on MIoU while BF1 gets a slight decrease. This shows that the removal of GRM has a greater impact on the overall segmentation and a certain loss in the boundary segmentation. It demonstrates the superiority of adopting GRB in the image overall segmentation problem.

● **Effectiveness of CGM:** in an effort to demonstrate the effectiveness of the deployed CGM, CGM is removed. The results are listed in the 3rd row of TABLE I. There is a decrease on MIoU and BF1. This might be because the pooling operations inside CGM enlarge the receptive field of the whole network compared to the light backbone, while the light backbone still needs to merge feature maps from different levels, indicating that our CGM is effective for solving the aliasing effect of upsampling. It also demonstrates the superiority of adopting CGM in the image overall segmentation problem.

● **Effectiveness of FFAB:** in an effort to demonstrate the effectiveness of the deployed FFAB, FFAB is substituted with concatenation and traditional convolutional layers. The results are listed in the 4th row of TABLE I. There is a significant decrease on MIoU and BF1. It shows that the fusion and amplification of effective information at different levels plays an important role in segmentation, which proves the effectiveness of FFAB.

2) **Loss ablation:** TABLE II and Fig.7 show the loss

TABLE III
SEGMENTATION PERFORMANCE OF DIFFERENT METHODS

| Method | Parameters | GFLOPS | Dataset | MIoU | PA | F1 | BF1 | | |
|---|---|---|---|---|---|---|---|---|---|
| | | | | | | | θ=3, θ'=3 | θ=3, θ'=5 | θ=3, θ'=7 |
| U-Net | 51.10M | 124.18 | Dataset1 | 0.9695 | 0.9948 | 0.9844 | 0.3230 | 0.3901 | 0.4242 |
| | | | Dataset2 | 0.9840 | 0.9934 | 0.9919 | 0.3999 | 0.4648 | 0.4917 |
| SegNet | 112.32M | 160.56 | Dataset1 | 0.9902 | 0.9984 | 0.9950 | 0.6806 | 0.7408 | 0.7664 |
| | | | Dataset2 | 0.9854 | 0.9940 | 0.9926 | 0.3972 | 0.4763 | 0.5124 |
| DFN | 470.78M | 615.27 | Dataset1 | **0.9911** | **0.9985** | **0.9955** | 0.6440 | 0.6906 | 0.7076 |
| | | | Dataset2 | 0.9858 | **0.9942** | 0.9928 | 0.3926 | 0.4542 | 0.4791 |
| ENet | 0.37M | 3.83 | Dataset1 | 0.9760 | 0.9960 | 0.9877 | 0.3122 | 0.3982 | 0.4443 |
| | | | Dataset2 | 0.9810 | 0.9922 | 0.9904 | 0.2868 | 0.3735 | 0.4242 |
| LEDNet | 3.50M | 5.72 | Dataset1 | 0.9834 | 0.9972 | 0.9916 | 0.3844 | 0.4442 | 0.4694 |
| | | | Dataset2 | 0.9832 | 0.9931 | 0.9915 | 0.3176 | 0.3728 | 0.3974 |
| Ours | 15.45M | 11.22 | Dataset1 | **0.9911** | **0.9985** | **0.9955** | **0.9391** | **0.9758** | **0.9860** |
| | | | Dataset2 | **0.9859** | **0.9942** | **0.9929** | **0.7996** | **0.9277** | **0.9592** |

ablation of hybrid loss based on the proposed BGHNet architecture. Obviously, the proposed hybrid loss is excellent in both quantitative evaluation and qualitative visual effect, especially for the boundary quality.

### D. Lightweight and Accuracy Analysis

In this section, the speed of proposed algorithm is firstly analyzed. Then, final results of proposed method on tongue datasets are compared with five of the most recent deep learning based methods: U-Net [51], SegNet [22], DFN [29], ENet [21], LEDNet [43].

**1) Lightweight analysis:** lightweight is a vital factor of an algorithm especially when we apply it in practice. Experiments were carried out in different settings by comparison. First, the status of FLOPS and parameters are shown in TABLE III. The FLOPS and parameters indicate the number of operations to process images of this resolution. For fair comparison, 512×512 is taken as the resolution of the input image. Meanwhile, the lightweight and corresponding accuracy results on tongue datasets are listed in TABLE III. From TABLE III, it can be found that our method achieves significant progress against the non-real-time semantic segmentation algorithms both in lightweight and accuracy. Compare with real-time semantic segmentation algorithms, although the lightweight is not as good enough, the accuracy is higher than them. In addition, the tongue boundaries of our results are better than both no-real-time and real-time algorithms.

**2) Accuracy analysis:** BGHNet can also achieve higher accuracy result against other real-time and no-real-time semantic segmentation algorithms, as shown in TABLE III. The quantitative and visual results for all compared methods on Dataset1 and Dataset2 are provided in TABLE III and Fig. 8, respectively. Our experimental images may cover most areas of the face, clothes, hair, some instruments and so on. However, in order to protect the privacy of patients' facial information and conveniently observe the experimental results, Fig. 8 only displays regions of the tongue and its surrounding areas in a limited range. The proposed method ranks the first place in the tongue segmentation with all evaluation indicators. Although it is tied with DFN on the overall segmented indicators MIoU, PA and F1, our model is far superior to DFN in terms of boundary accuracy and model lightweight. Especially, our network outperforms most of the other methods by a large margin. From Fig. 8, it can be seen that the visual segmentation results of BGHNet is much better than other methods. It demonstrates the advantages of the proposed method in dealing with the challenges of the tongue segmentation.

## IV. CONCLUSION

Bilateral Guidance Hierarchical Network (BGHNet) is proposed in this paper to improve the lightweight and accuracy of real-time semantic tongue image segmentation. The proposed BGHNet is a hierarchical and progressive predict-refine architecture. It consists of two components: a Context Feature Encoder Module (CFEM) and a novel Hierarchical Feature Refinement Module (HFRM). Both components are lightweight that fast training and deployment are possible. Combined with a hybrid loss, BGHNet can not only capture object structures of different scales and fine object structures, but also produce object maps with clear boundaries. Experimental results on two tongue datasets demonstrate that our model outperforms other methods in terms of both region-based and boundary-aware measures. The proposed method can be employed to solve complicated medical image analysis problems, even with limited training data. Additionally, our proposed network architecture is modular. Hence it can be flexibly and easily extended or adapted to other tasks by replacing either the predicting network or the refinement module.

## REFERENCES

[1] M. H. Tania, K. Lwin and M. A. Hossain, "Advances in automated tongue diagnosis techniques," Integrative Medicine Research., vol.8, no.1, pp. 42-56, March. 2019.
[2] G. Maciocia, "Tongue diagnosis in Chinese medicine," Seattle, WA: Eastland, 1995.
[3] D. Zhang, H. Zhang and B. Zhang, "Tongue image analysis for appendicitis diagnosis," Tongue Image Analysis., Singapore, vol. 175, no.3, pp. 160-176, 14. October. 2005.
[4] L. Xu, Y. Sun and B. Wang, "Application value of tongue diagnosis in health management of stroke based on literature," Chinese Journal of Basic Medicine in Traditional Chinese Medicine., vol. 25, no. 3, pp. 326-329, 2019.
[5] J. Yang, Y. Hong and S. Ma, "Impact of the new health care reform on hospital expenditure in China: A case study from a pilot city," China Econ. Rev., vol. 39, pp. 1–14, Jul. 2016.
[6] C. C. Yang and P. Veltri, "Intelligent healthcare informatics in big data era," Artif. Intell. Med., vol. 65, no. 2, pp. 75–77, 2015.
[7] P. Chen, L. Xiao, X. Zhang and P. Feng, "Telehealth attitudes and use among medical professionals, medical students and patients in China: A cross-sectional survey," Int. J. Med. Informat., vol. 108, pp. 13–21, Dec. 2017.
[8] M. Zieba, "Service-oriented medical system for supporting decisions with missing and imbalanced data," IEEE J. Biomed. Health Informat., vol. 18, no. 5, pp. 1533–1540, Sep. 2014.
[9] R. K. Lomotey, J. Nilson, K. Mulder, K. Wittmeier, C. Schachter and R. Deters, "Mobile medical data synchronization on cloud-powered middleware

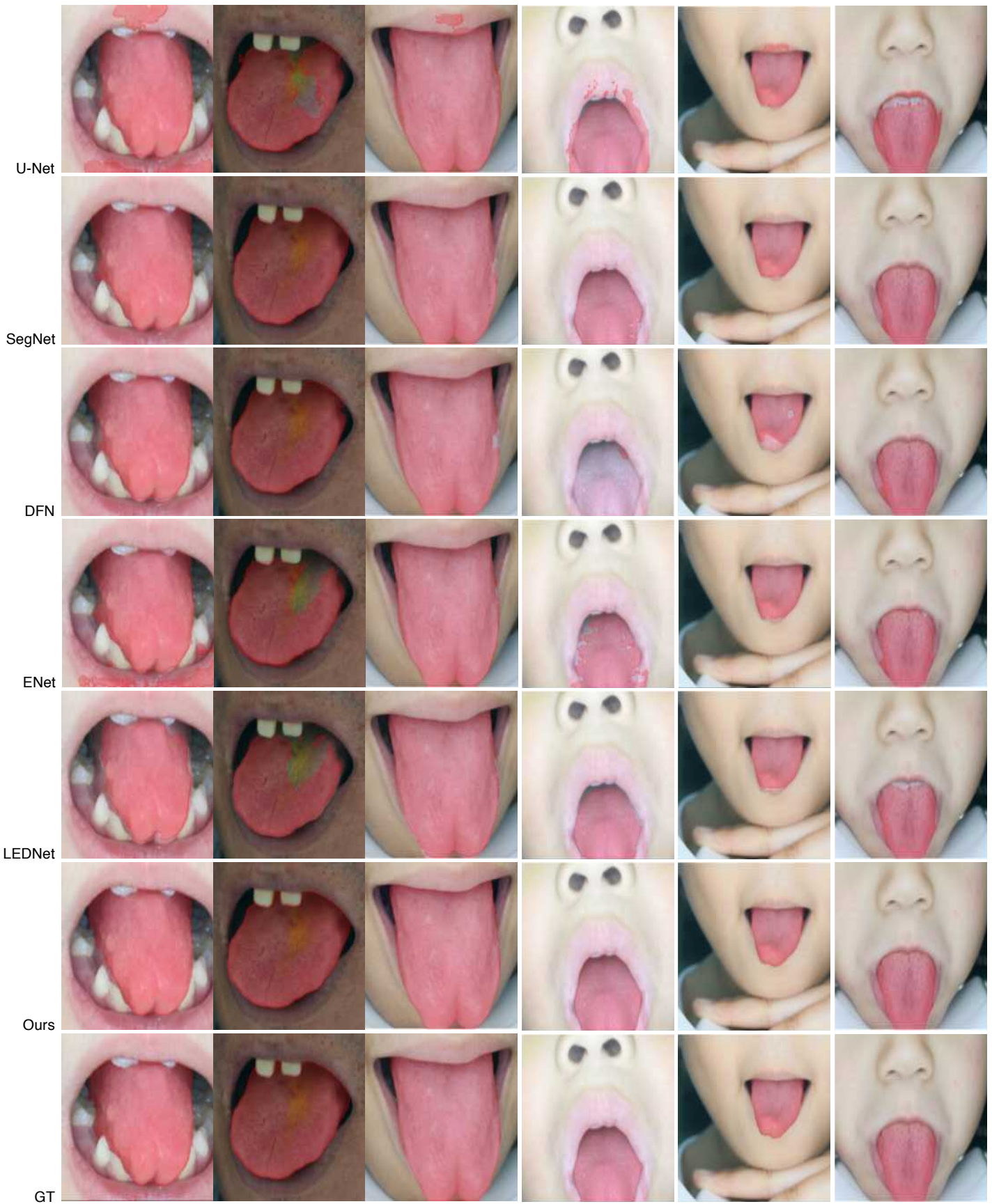

Fig. 8. Sample results of different methods on Dataset1. From top to bottom: state-of-the-art results obtain by U-Net [37], SegNet [22], DFN [29], ENet [21], LEDNet [43], BGHNet and ground-truth masks (Red transparent area indicates the mask as the segmentation prediction map or ground truth).

platform," in IEEE Transactions on Services Computing, vol. 9, no. 5, pp. 757-770, 1 Sept.-Oct. 2016.
[10] C. Dainton and C. H. Chu, "A review of electronic medical record keeping on mobile medical service trips in austere settings," Int. J. Med. Informat., vol. 98, pp. 33–40, Feb. 2017.
[11] X. Zhai, H. Lu and L. Zhang, "Application of image segmentation technique in tongue diagnosis," 2009 International Forum on Information Technology and Applications, Chengdu, 2009, pp. 768-771.
[12] J. Ning, D. Zhang, C. Wu and Y. Feng, "Automatic tongue image segmentation based on gradient vector flow and region merging," Neural Computing and Applications., vol. 21, no. 8, pp. 1819-1826, 2012.
[13] D. S. Yang, Y. H. Chen, F. M. Zou and C. L. Zhou, "Effective algorithm on automatic segmentation of body of tongue," Application Research of Computers., vol. 24, no. 2, pp. 170-172, 2007.
[14] Z. C. Yu, Z. C. Zhang, Z. Y. Li and W. N. Liu, "Threshold segmentation algorithm of tongue image based on multi-color components," Computer Applications and Software., vol. 36, no. 5, 2019.
[15] Z. C. Fu, X. Q. Li and F. F. Li, "Tongue image segmentation based on snake model and radial edge detection," Journal of Image and Graphics., vol. 14, no. 4, pp. 688-693, 2018.
[16] Z. S. Zhang, "Improvement wavelet transformation in Chinese medicine pip edge examination research," Computer Engineering and Applications., vol. 48, no. 35, pp. 135-138, 2012.
[17] J. Ning, L. Zhang, D. Zhang and C. Wu, "Interactive image segmentation by maximal similarity based region merging," Pattern Recognit., vol. 43, pp. 445-456, February 2010.
[18] M. Y. Wang, X. F. Zhang and L. Zhuo, "An improved snaked model for tongue image segmentation," Measurement & Control Technology., vol. 30, no. 5, pp. 32-36, 2011.
[19] C. Zhou, H. Fan and Z. Li, "Tonguenet: accurate localization and segmentation for tongue images using deep neural networks," in IEEE Access, vol. 7, pp. 148779-148789, 2019.
[20] J. Zhou, Q. Zhang, B. Zhang and X. Chen, "TongueNet: a precise and fast tongue segmentation system using U-Net with a morphological processing layer," Applied Sciences., vol. 15, no. 9, pp. 3128, 2019.
[21] P A. Paszke, A. Chaurasia, S. Kim and E.Culurciello. (Jun.2016). "Enet: a deep neural network architecture for real-time semantic segmentation." [Online]. Available: https://arxiv.org/abs/1606.02147
[22] V. Badrinarayanan, A. Kendall and R. Cipolla, "SegNet: a deep convolutional encoder-decoder architecture for image segmentation," in IEEE Transactions on Pattern Analysis and Machine Intelligence, vol. 39, no. 12, pp. 2481-2495, 1 Dec. 2017.
[23] H. Zhao, J. Shi, X. Qi, X. Wang and J. Jia, "Pyramid scene parsing network," 2017 IEEE Conference on Computer Vision and Pattern Recognition (CVPR), Honolulu, HI, 2017, pp. 6230-6239.
[24] L. Chen, G. Papandreou, I. Kokkinos, K. Murphy and L. A. Yuille. (Dec.2014). "Semantic image segmentation with deep convolutional nets and fully connected crfs." [Online]. Available: https://arxiv.org/abs/1412.7062
[25] L. Chen, G. Papandreou, I. Kokkinos, K. Murphy and A. L. Yuille, "DeepLab: semantic image segmentation with deep convolutional nets, atrous convolution, and fully connected CRFs," in IEEE Transactions on Pattern Analysis and Machine Intelligence, vol. 40, no. 4, pp. 834-848, 1 April 2018.
[26] G. Bertasius, J. Shi and L. Torresani, "Semantic segmentation with boundary neural fields," 2016 IEEE Conference on Computer Vision and Pattern Recognition (CVPR), Las Vegas, NV, 2016, pp. 3602-3610.
[27] Z. Yu, C. Feng, M. Liu and S. Ramalingam, "CASENet: deep category-aware semantic edge detection," 2017 IEEE Conference on Computer Vision and Pattern Recognition (CVPR), Honolulu, HI, 2017, pp. 1761-1770.
[28] B. Hariharan, P. Arbeláez, L. Bourdev, S. Maji and J. Malik, "Semantic contours from inverse detectors," 2011 International Conference on Computer Vision, Barcelona, 2011, pp. 991-998.
[29] C. Yu, J. Wang, C. Peng, C. Gao, G. Yu and N. Sang, "Learning a discriminative feature network for semantic segmentation," 2018 IEEE/CVF Conference on Computer Vision and Pattern Recognition, Salt Lake City, UT, 2018, pp. 1857-1866.
[30] Z. Wang, E. P. Simoncelli and A. C. Bovik, "Multiscale structural similarity for image quality assessment," The Thrity-Seventh Asilomar Conference on Signals, Systems & Computers, 2003, Pacific Grove, CA, USA, 2003, pp. 1398-1402 Vol.2.
[31] G. Csurka, D. Larlus and F. Perronnin, "What is a good evaluation measure for semantic segmentation?," BMVC, 2013, 27: 2013.
[32] A. VBokhovkin and E. Burnaev, "Boundary loss for remote sensing imagery semantic segmentation," 2019 International Symposium on Neural Networks, Springer, Cham, pp. 388-401.
[33] W. Luo, Y. Li, R. Urtasun and R. Zemel, "Understanding the effective receptive field in deep convolutional neural networks," in Advances in neural information processing systems, 2016, pp. 4898-4906.
[34] J. Hu, L. Shen and G. Sun, "Squeeze-and-excitation networks," 2018 IEEE/CVF Conference on Computer Vision and Pattern Recognition, Salt Lake City, UT, 2018, pp. 7132-7141.
[35] C. Yu, J. Wang, C. Peng, C. Gao, G. Yu and N. Sang, "Bisenet: bilateral segmentation network for real-time semantic segmentation," in ECCV, 2018, pp. 325-341.
[36] P. T. De Boer, D. P. Kroese, S. Mannor and R. Y. Rubinstein, "Atutorial on the cross-entropy method," Annals OR, vol. 134, no. 1, pp. 19-67, 2005.
[37] O. Ronneberger, P. Fischer and T. Brox, "U-net: convolutional networks for biomedical image segmentation," in International Conference on Medical image computing and computer-assisted intervention, 2015, pp. 234-241.
[38] W. Shi, J. Caballero, F. Huszár, J. Totz, P. A. Aitken, R. Bishop, et al., "Real-time single image and video superresolution using an efficient sub-pixel convolutional neural network," in Proceedings of the IEEE conference on computer vision and pattern recognition, 2016, pp. 1874-1883.
[39] H. Noh, S. Hong and B. Han, "Learning deconvolution network for semantic segmentation," 2015 IEEE International Conference on Computer Vision (ICCV), Santiago, 2015, pp. 1520-1528.
[40] L. C. Chen, Y. K. Zhu, G. Papandreou, F. Schroff, and H. Adam. (Feb.2018). "Encoder-decoder with atrous separable convolution for semantic image segmentation." [Online] Available: https://arxiv.org/abs/1802.02611
[41] F. Chollet, "Xception: deep learning with depthwise separable convolutions," 2017 IEEE Conference on Computer Vision and Pattern Recognition (CVPR), Honolulu, HI, 2017, pp. 1800-1807.
[42] X. Qin, Z. Zhang, C. Huang, C. Gao, M. Dehghan and M. Jagersand, "BASNet: boundary-aware salient object detection," 2019 IEEE/CVF Conference on Computer Vision and Pattern Recognition (CVPR), Long Beach, CA, USA, 2019, pp. 7471-7481.
[43] Y. Wang, Q. Zhou, J. Liu, J. Xiong, G. Gao, X. Wu, et al., "Lednet: a lightweight encoder-decoder network for real-time semantic segmentation," 2019 IEEE International Conference on Image Processing (ICIP), Taipei, Taiwan, 2019, pp. 1860-1864.
[44] F. Yu and V. Koltun, "Multi-scale context aggregation by dilated convolutions," in ICLR, 2016.
[45] Z. Wang, E. P. Simoncelli and A. C. Bovik, "Multiscale structural similarity for image quality assessment," The Thrity-Seventh Asilomar Conference on Signals, Systems & Computers, 2003, Pacific Grove, CA, USA, 2003, pp. 1398-1402 Vol.2.
[46] A. Krizhevsky, I. Sutskever and G. E. Hinton, "Imagenet classification with deep convolutional neural networks," in Advances in neural information processing systems, 2012, pp. 1097-1105.
[47] W. Liu, A. Rabinovich and A. C. Berg. (Jun.2015). "Parsenet: looking wider to see better." [Online] Available: https://arxiv.org/abs/1506.04579
[48] H. Chen, X. Qi, L. Yu and P. Heng, "DCAN: deep contour-aware networks for accurate gland segmentation," 2016 IEEE Conference on Computer Vision and Pattern Recognition (CVPR), Las Vegas, NV, 2016, pp. 2487-2496.
[49] B. Lin, J. Xie, C. Li and Y. Qu, "Deeptongue: tongue segmentation via resnet," 2018 IEEE International Conference on Acoustics, Speech and Signal Processing (ICASSP), Calgary, AB, 2018, pp. 1035-1039.